\documentclass[authoryear,final,12pt]{elsmodarticle}

\usepackage{amsmath,amsfonts,amsthm,amssymb}
\usepackage{graphicx}
\usepackage{color}

\newtheorem {thm}     {Theorem}
\newtheorem {lem} {Lemma}

\newtheorem {deff} {Definition}

\newtheorem{prop}{Proposition}

\newcommand{\Su}{\textbf{S}}
\newcommand{\I}{\textbf{I}}
\newcommand{\R}{\textbf{R}}
\newcommand{\Nt}{{(N,\Delta t)}}
\newcommand{\bx}{\mathbf{x}}
\newcommand{\by}{\mathbf{y}}
\newcommand{\rd}{\mathrm{d}}
\newcommand{\re}{\mathrm{e}}
\newcommand{\supp}{\mathrm{supp}}
\newcommand{\bydef}{:=}

\newcommand{\Si}{\mathcal{S}}

\renewcommand{\div}{\mathrm{div}}

\begin{document}

\begin{frontmatter}

\title{The SIR epidemic model from a PDE point of view}

\author{Fabio A. C. C. Chalub}
\address{Departamento de Matem\'atica and Centro de Matem\'atica
e Aplica\c c\~oes, Universidade Nova de Lisboa, 
Quinta da Torre, 2829-516, Caparica, Portugal.}
\ead{chalub@fct.unl.pt}
\author{Max O. Souza}
\address{Departamento de Matem\'atica Aplicada, Universidade Federal
Fluminense, R. M\'ario Santos Braga, s/n, 22240-920, Niter\'oi, RJ, Brasil.}
\ead{msouza@mat.uff.br}

\date{\today}

\begin{abstract}
We present a derivation of the classical SIR model through a mean-field approximation from a discrete version of SIR. We then obtain a hyperbolic forward Kolmogorov equation, and show that its projected characteristics recover the standard SIR model. Moreover, we show that the long time limit of the evolution will be a Dirac measure. 
The exact position will depend on the well-know $R_0$ parameter, and it will be supported on the corresponding stable SIR equilibrium.
\end{abstract}

\end{frontmatter}

\section{Introduction}

A very fruitful modeling paradigm in epidemiology is the so-called compartmental models, with dynamics governed by mass-action laws. Most classical epidemiological models are of this type, and this has led to a number of  both quantitative and qualitative predictions in the disease dynamics~\cite{Britton}. 
More recently, there is a growing interest in discrete, agent-based, models~\cite{Pacheco}. 
See also~\cite{Schneckerreithera} for a comparison between different models.
In many cases, these models are thought to be more realistic, and  able to capture important dynamical features that are not present in the continuous models. 

Here, we follow the ideas in \cite{ChalubSouza09b}, to study the the large-population regime of the discrete dynamics. In this way, we obtain a Hyperbolic Forward Kolmogorov equation for the probability density evolution. It is well known, through  the method of characteristics, that there is a strong linkage between solutions to first-order Partial Differential Equations (PDEs) and systems of Ordinary Differential Equations (ODEs). Thus, its not entirely surprising that projected characteristics of this PDE will be related to the classical SIR (Susceptible-Infected-Removed) ODE model.

This modeling through PDEs has some advantages, in particular, it allows the introduction of higher-order effects,
like, for example, stochasticity, adding a second-order term to the equation. On the other hand, if we consider 
a discrete model in population dynamics and consider its limit of large population (under suitable 
condition) we naturally obtain a PDE for $p(t,x)$, the
probability density to find the population at state $x$ at time $t$. The ODE can then be obtained as 
the hyperbolic limit of the PDE, or, alternatively as the initial dynamics of the PDE. In short, this
means that the dynamics of the discrete population can be approximated for short times and large population
by a certain ODE --- the derivation of this ODE requires an introduction of intermediate models, 
a stochastic differential equation or a partial differential one.

The ODE approach can be seen in~\cite{Allen}, while the PDE modeling was the subject of a previous work from the authors, where the replicator equation was obtained as the limit of the finite-population discrete Moran 
process~\cite{ChalubSouza09b}. The resulting equation is of singular type and required a specific analysis of its behavior~\cite{ChalubSouza09}.

In this work, we will study in a certain level of detail the SIR epidemic
model from the PDE point of view\footnote{The expression SIR appears in the literature in two different context:
one as a discrete evolutionary system, used in general in computer simulations; the second as an ODE system. We hope
that all these different meanings are clear from the context.}. This is one of the most elementary and well studied model in 
mathematical epidemiology. See~\cite{Murray_vol1,Britton}. 


In particular, we shall prove that the solution of the SIR-hyperbolic-PDE obtained as a first order expansion in the inverse of
the population size from the discrete-SIR converges when $t\to\infty$ to a Dirac-delta measure supported at
the unique stable equilibrium of the SIR-ODE. Moreover (and this will be proved in a forthcoming work)
the solution of the SIR-hyperbolic is approximated by the SIR-ODE in all time scales and approximate the
discrete-SIR for short time scales. This provides a framework to unify all these descriptions. In a 
forthcoming work, we will also introduce the SIR-parabolic-PDE which has the inverse behavior (approximate the 
discrete-SIR for all time scales and the ODE-SIR for short times).
 

\section{Discrete and continuous SIR models}

Consider a discrete SIR model, i.e., consider a fixed size population of $N$ individuals, each one
in one of the three states: $n$ individuals \textbf{S}usceptible (i.e, individuals with no imunity), $m$ individuals \textbf{I}nfected (i.e., individual currenctly infected by a given infectious disease and able to transmit it to
the susceptibles) 
and $N-n-m$ \textbf{R}emoved (after infection idividuals have a temporary imunity and then are removed
from the dynamics; after certain time they become susceptible again). This is a very simple model for
non lethal diseases transmitted by contact, e.g., normal influenza. 
At each time step of size $\Delta t>0$ we select one individual at random:
\begin{itemize}
\item If it is \Su, then it changes to \I\ with probability proportional to the fraction of \I\ in the
remainder, $\alpha m/(N-1)$;
\item If it is \I, then it changes to \R\ with constant probability $\beta$;
\item If it is \R, then it changes to \Su\ with constant probability $\gamma$.
\end{itemize}

This can be summarized in the following diagram:
\begin{equation*}
\mathbf{S}+\mathbf{I}\stackrel{\alpha}{\longrightarrow}\mathbf{I}+\mathbf{I}\ ,\qquad
\mathbf{I}\stackrel{\beta}{\longrightarrow}\mathbf{R}\ ,\qquad
\mathbf{R}\stackrel{\gamma}{\longrightarrow}\mathbf{S}\ .
\end{equation*}

This model is also called SIRS, and when $\gamma=0$ we recover the classical SIR model.
For simplicity, however, we will call it the SIR model in this work and all results 
presented here include the case $\gamma=0$.


Constants $\alpha$, $\beta$ and $\gamma$ depend, in principle, in $N$ and $\Delta t$. As we are
interested in the limit behavior when $N\to\infty$, $\Delta t\to 0$ we will assume the
following \textit{scaling} relations
\[
\lim_{N\to\infty,\Delta t\to 0}\frac{\alpha}{N\Delta t}=a\ ,\qquad
\lim_{N\to\infty,\Delta t\to 0}\frac{\beta}{N\Delta t}=b\ ,\qquad
\lim_{N\to\infty,\Delta t\to 0}\frac{\gamma}{N\Delta t}=c\ ,
\]
with $ab\ne0$.
For further informations on scalings, see~\cite{ChalubSouza09b}.

Let $P_\Nt(t,n,m)$ be the probability that at time $t$ we have
$n$ susceptible, $m$ infected and $N-n-m$ removed, where the
total population $N$ is constant and the time step is given by $\Delta t>0$. Therefore
\begin{eqnarray*}
&&P_\Nt(t+\Delta t,n,m)=\alpha \frac{(n+1)(m-1)}{N(N-1)} P_\Nt(t,n+1,m-1)\\
&&\qquad+\beta \frac{m+1}{N}P_\Nt(t,n,m+1)+\gamma \frac{N-n-m+1}{N}P_\Nt(t,n-1,m)\\
&&\qquad+\left[\frac{n}{N}\left(1-\alpha\frac{m}{N-1}\right)+\frac{m}{N}(1-\beta)+
\frac{N-n-m}{N}(1-\gamma)\right]P_\Nt(t,n,m)\ .
\end{eqnarray*}

Now, define $x=n/N$, $y=m/N$ and $p(t,x,y)=P(t,xN,yN;N)$. Then, using $p(t,x,y)=p$
and keeping terms until order $1/N$:
\begin{eqnarray*}
&&p(t+\Delta t,x,y)=\alpha\frac{\left(x+\frac{1}{N}\right)\left(y-\frac{1}{N}\right)}
{\left(1-\frac{1}{N}\right)}p\left(t,x+\frac{1}{N},y-\frac{1}{N}\right)\\
&&\qquad+\beta\left(y+\frac{1}{N}\right)p\left(t,x,y+\frac{1}{N}\right)+
\gamma\left(1-x-y+\frac{1}{N}\right)p\left(t,x-\frac{1}{N},y\right)\\
&&\qquad +\left(x\left(1-\frac{\alpha y}{1-\frac{1}{N}}\right)+y(1-\beta)+(1-x-y)(1-\gamma)\right)
p(t,x,y)\\
&&\approx
p+\frac{1}{N}\left[\left(\alpha\left(y-x\right)+\beta+\gamma\right)p
+\left(\alpha x y-\gamma(1-x-y)\right)\partial_x p+\left(\beta y-\alpha xy\right)\partial_y p
\right]\\
&&=p+
\frac{1}{N}\left[\partial_x\left((\alpha xy-\gamma(1-x-y))p\right)+
\partial_y\left((\beta-\alpha x)yp\right)\right]
\end{eqnarray*}
Finally,
\begin{equation}\label{SIRPDE1}
 \partial_t p=\partial_x\left((axy-c(1-x-y))p\right)+\partial_y\left((b-ax)yp\right)\ .
\end{equation}
subject to probability conservation, i.e, 
\begin{equation}
 \frac{\rd}{\rd t}\int_0^1p(t,x)\rd x =0.\label{consprob}
\end{equation}
%


\section{The SIR model as a transport problem}

Let $\bx=(x,y)$, and let $\Phi_t(\bx)$ be the flow map associated to the SIR system
\begin{eqnarray*}
\dot X&=&c(1-X-Y)-a XY\ ,\\
\dot Y&=&(aX-b)Y\ ,
\end{eqnarray*}
then, if $p_0(\bx)$ is $C^1(R^2)$ function, and $p(t,x,y)=\re^{Q(\bx)-Q(\Phi_{-t}(\bx))}p_0(\Phi_{-t}(\bx))$, with $\bx=(x,y)$, and $Q$ satisfies
%
%
\[
 F\cdot\nabla Q = - \nabla\cdot F,
\]
where $F$ denotes the right hand side of the SIR system. Let $\Si$ denote the unit simplex in $\mathbb{R}^2$. 

Fix $\bx_0\in\Si$. Then, we have that 
\[
 \re^{-Q(\phi_t(\bx_0))}p(t,\phi_t(\bx_0))=\re^{Q(\bx_0)}p_0(\bx_0).
\]
Hence,
\begin{align*}
0&=\re^{Q(\phi_t(\bx_0))}\frac{\rd }{\rd t}\left[ \re^{-Q(\phi_t(\bx_0))}p(t,\phi_t(\bx_0))\right]\\
&= -F(\phi_t(\bx_0))\cdot\nabla Q(\phi_t(\bx_0)p(t,\phi_t(\bx_0))+\partial_tp(t,\phi_t(\bx_0)+F(\phi_t(\bx_0))\nabla p(t,\phi_t(\bx_0))\\
&=\nabla\cdot Fp + F\nabla p +\partial_tp\\
&=\partial_tp + \nabla\cdot\left(pF\right).
\end{align*}
Thus, the SIR system are the characteristics of \eqref{SIRPDE1}, and the probability density should be transported along them. We now make this calculation more precise. 

 \begin{deff}
  Let $F:U\subset\mathbb{R}^n\to\mathbb{R}^n$, be a Lipschitz vector field, where $U$ is an open set, and let $\Omega\subset U$ be  compact. We say that  $\Omega$ is regularly attracting for $F$, if there is an open set $V$ with a piecewise smooth boundary and $\Omega\subset V\subset \bar{V}\subset U$, such that, if $\Phi_t$ denotes the flow by $F$ restricted to $V$, then we have that $\omega(V)\subset\bar{\Omega}$. 
 \end{deff}
\begin{thm}\label{propsmooth}
 Let $\Omega$ be a domain with piecewise smooth boundary $\partial\Omega$. Let $F:U\subset\mathbb{R}^N\to\mathbb{R}^N$ be Lipschitz, with $\Omega\subset U$ being a regularly attracting set for $F$. Let $p_0\in L^1(\Omega)$ be nonnegative. Then the equation
\begin{equation}
 \partial_t p + \nabla\cdot\left(pF\right)=0,\quad p(0,x)=p_0(x)\label{modproblem}
\end{equation}
has a unique solution
\[
 p(t,\bx)=\re^{Q(\bx)-Q(\Phi_{-t}(\bx))}p_0(\Phi{-t}(\bx))
\]
Moreover, $p$ is nonnegative, and $\supp(p(t,\cdot))\subset\Omega$. 

In addition, if $\supp(p_0)\subset\Omega$, then
\begin{equation}
 \frac{\rd}{\rd t}\int_{\Omega}p(t,\bx)\rd\bx = 0. \label{modcons}
\end{equation}
\end{thm}
\begin{proof}
 Existence can be shown as follows by considering the weak formulation, Let $W=[0,\infty)\times\Omega$, and let $\psi\in C_c(W)$. 
\begin{align}
 &\iint_{W}p(t,\bx)\partial_t\psi(t,\bx)\rd\bx\rd t + 
\iint_{W}p(t,\bx)\nabla\psi(t,\bx)\rd\bx\rd t \quad + \nonumber\\
&\qquad + \int_{\Omega}p(0,\bx)\psi(0,\bx)\rd\bx\rd t=0.\label{weakform}
\end{align}
Choose $\psi\in C_c((0,\infty)\times\Omega)$. 
Then \eqref{weakform} becomes
\[
 \iint_{W}p(t,\bx)\partial_t\psi(t,\bx)\rd\bx\rd t + 
\iint_{W}p(t,\bx)\nabla\psi(t,\bx)\rd\bx\rd t \quad =0.
\]
Let $\by=\phi_t(x)$ and $\eta_t(x)=\mathrm{det}(\partial_x\Phi_t(x))$. Also let $W_t=\Phi_t(W)$.
Then, the first integral becomes
\[
 \iint_{W_t}\re^{Q(\Phi_t(\by))-Q(\by)}p_0(\by)\partial_t\psi(t,\Phi_t(y))\eta_t(\by)\rd\by\rd t.
\]
The second integral becomes
\[
 \iint_{W_t}\re^{Q(\Phi_t(\by))-Q(\by)}p_0(\by)\nabla\cdot\psi(t,\Phi_t(y))\eta_t(\by)\rd\by\rd t.
\]
Combining both integrals, we can write
\[
 \iint_{W_t}\re^{Q(\Phi_t(\by))-Q(\by)}p_0(\by)\frac{\rd}{\rd t}\psi(t,\Phi_t(\by))\eta_t(\by)\rd\by\rd t.
\]
On integrating by parts, we have that
\[
 \iint_{W_t}\re^{-Q(\by)}p_0(\by)\psi(t,\Phi_t(y))\re^{Q(\Phi_t(\by))}\left[F\cdot\nabla Q - \nabla\cdot F\right]\rd\by\rd t =0.
\]
We have that $p$ is clearly nonnegative. Moreover,
Let $V=\supp(u(t,\cdot))$ and let $V_t=(\Phi_{-t}(W))$. Let $\psi$ be a vanishing function in $\Omega$. Then
\[
 0=\int_{\Phi_t(\Omega)}u(t,\bx)\psi(\bx)\rd\bx=\int_{\Omega}\re^{Q(\Phi_{t}(\by)-Q(\by)}u_0(\by)\psi(\by)\div(F)\rd\by
\]
Hence $\supp(u(t,\cdot))\subset\Phi_t(\Omega) $. Since $\Omega$ is regularly attracting for $F$, we have that $\Phi_t(\Omega)\subset\Omega$.
If $\supp(p_0(\by))\subset\Omega$, then we extend $p_0$ by defining it to be zero in $\mathbb{R}^N-U$. Then
we have that 
\begin{align*}
 \frac{\rd}{\rd t}\int_{\Omega}u(t,\cdot)\rd\bx&=\frac{\rd}{\rd t}\int_{V}u(t,\cdot)\rd\bx\\
&=-\int_{V}\div(u(t,\cdot)\hat{F})\rd\bx=0.
\end{align*}
\end{proof}

%
%
%
%
%
%
%
%
%
%
%
%
Let
\[
 F_S(x,y)=(c(1-x-y)-\alpha xy,\alpha xy - by).
\]
We define the reflected SIR field by
\[
 F_{RS}(x,y)=(c(1-x-y)-a xy,(a x - b)|y|)
\]
We immediately have
\begin{lem}
 Let $\Omega$ be the simplex in the nonnegative orthant of $R^2$.  Then $F_{RS}$ is a $C^1$ vector field in $\mathbb{R}^2$, and $\Omega$ is regularly attracting for $F_{RS}$.
\end{lem}
Thus for smooth solutions, we have 
\begin{thm}
Consider the Cauchy problem for \eqref{SIRPDE1}, with a $L^1$ non-negative initial condition $p_0$. Then there exists a unique solution satisfying \eqref{consprob}. Moreover, $p(t,x)\geq0$.
\end{thm}

\section{Asymptotic Behavior and measure solutions}

We recall that the dynamics of SIR are controlled by the parameter $R_0\bydef a/b$ (see~\cite{Britton}), i.e,
\begin{prop}
 Let $\bx_1=(1,0)$, and $\bx_2=\left(\frac{1}{R_0},\frac{c}{c+b}\left(1-\frac{1}{R_0}\right)\right)$ be the two equilibria of SIR. $\bx_1$ is referred to as the \textit{disease free equilibrium} and $\bx_2$ as the \textit{endemic equilibrium}. If $R_0\leq 1$, then any solution that starts in the nonnegative orthant of $\mathbb{R}^2$ approaches $\bx_0$ for large time. If $R_0>1$, then $\bx_1$ is the limiting point.
\end{prop}

With this point of view we have
\begin{thm}
 Let $p$ be a solution of \eqref{SIRPDE1}, satisfying \eqref{consprob}. Then, in the Wasserstein metric, we have that
\[
 \lim_{t\to\infty}p(t,x)=\left\{
\begin{array}{lr}
 \delta_{\bx_1},&R_0\leq1\\
\delta_{\bx_2},&R_0>1.
\end{array}\right.
\]
\end{thm}
\begin{proof}
 We deal with the case $R_0\leq 1$; the case $R_0>1$ is analogous. Since $R_0\leq1$, we have that $\bx_1$ is the globally asymptotic stable equilibrium. Then given, $\delta>0$, we can find $T>0$, such that, for $t>T$, we have that
\[
 \Phi_t(\Si)\subset B_{\delta}(\bx_1).
\]
Let $\psi(\bx)$ be a continuous function.
Then, for $t>T$, we have that
\[
\int_{\Si}p(t,\bx)\psi(\bx)\rd\bx = \int_{B_\delta(\bx_1)}p(t,\bx)\psi(\bx)\rd\bx.
\]
But, let $\epsilon>0$ be given. Since $\psi$ is continuous, we have $\delta>0$, such that
\[
 \psi_(\bx_1)-\epsilon\leq\int_{B_\delta(\bx_1)}p(t,\bx)\psi(\bx)\rd\bx\leq\psi(\bx_0)+\epsilon,
\]
and this proves the claim, since we have that
\begin{align*}
 \int_{B_\epsilon(x_1)}p(t,\Phi_t(\by)\rd\by&= 
\int_{B_\epsilon(x_1)}\re^{Q(\Phi_t(\by))-Q(\by))}p_0(\by)\psi(\Phi_t(\by))\eta_t(\by)\rd y\\
&
\end{align*}

\end{proof}
%

%

\section*{Acknowledgments}

FACCC is partially supported by FCT/Portugal, grants  
 PTDC/MAT/68615/2006 and PTDC/MAT/66426/2006, PTDC/FIS/7093/2006. 
MOS is partially supported by FAPERJ grants  170.382/2006 and 110.174/2009. Both FACC and MOS are partially supported by the bilateral agreement Brazil-Portugal (CAPES-FCT). 

\bibliography{SIRPDE}

\begin{thebibliography}{1}

\bibitem{Allen}
Linda J.~S. Allen.
\newblock An introduction to stochastic epidemic models.
\newblock In {\em Mathematical epidemiology}, volume 1945 of {\em Lecture Notes
  in Math.}, pages 81--130. Springer, Berlin, 2008.

\bibitem{Britton}
Nicholas~F. Britton.
\newblock {\em Essential mathematical biology}.
\newblock Springer Undergraduate Mathematics Series. Springer-Verlag London
  Ltd., London, 2003.

\bibitem{ChalubSouza09b}
F.A.C.C Chalub and M.~O. Souza.
\newblock From discrete to continuous evolution models: A unifying approach to
  drift-diffusion and replicator dynamics.
\newblock {\em Theoretical Population Biology}, 76(4):268--277, 2009.

\bibitem{ChalubSouza09}
F.A.C.C Chalub and M.~O. Souza.
\newblock A non-standard evolution problem arising in population genetics.
\newblock {\em Commun. Math. Sci.}, 7(2):489--502, 2009.

\bibitem{Murray_vol1}
J.~D. Murray.
\newblock {\em Mathematical biology. {I}}, volume~17 of {\em Interdisciplinary
  Applied Mathematics}.
\newblock Springer-Verlag, New York, third edition, 2002.
\newblock An introduction.

\bibitem{Schneckerreithera}
G.~Schneckenreithera, N.~Poppera, G.~Zaunera, and F.~Breiteneckera.
\newblock Modelling sir-type epidemics by {ODE}s, {PDE}s, difference equations
  and cellular automata – a comparative study.
\newblock {\em Simulation Modelling Practice and Theory}, 16(8):1014--1023,
  2008.

\bibitem{Pacheco}
J.~Verdasca, M.~M. Telo~da Gama, A~Nunes, N.~R. Bernardino, J.~M. Pacheco, and
  M.~C. Gomes.
\newblock Recurrent epidemics in small world networks.
\newblock {\em J. Theor. Biol.}, 233:553--561, 2005.

\end{thebibliography}
\bibliographystyle{plain}



\end{document}